\documentclass[11pt]{article}
\usepackage{mathrsfs}
\usepackage{tikz}
\usepackage{verbatim}
\usepackage{amsmath}\numberwithin{equation}{section}
\usepackage{amssymb}
\usepackage{amsfonts}
\usepackage{CJK}
\usepackage{subfigure}
\usepackage{graphicx}
\usepackage{dsfont}
\usepackage{amscd}%\usepackage{showkeys}
\usepackage[body={16.5cm,26cm}, top=1cm]{geometry}
\newtheorem{thm}{Theorem}

\usetikzlibrary{arrows}
\begin{document}

\title{Reply to ``Comment on ``Darboux transformation and
classification of solution for mixed coupled
nonlinear Schr\"odinger equations""}
\author{Liming Ling$^1$, Li-Chen Zhao$^2$, Boling Guo$^3$\\
$^1$School of Mathematics, South China University of Technology, Guangzhou 510640, China;\\ Email: linglm@scut.edu.cn\\
$^2$Department of Physics, Northwest University, 710069, Xi'an, China;\\
$^3$Institute of Applied Physics and Computational Mathematics, 100088, Beijing, China\\}
\maketitle

\begin{abstract}
In the recent comment quoted in the title (arXiv:1407.7852v1), a comment is presented on our recent work which derive a generalized nonlinear wave solution formula for mixed coupled nonlinear Sch\"{o}dinger equations by performing the unified Darboux transformation  (arXiv:1407.5194).  Here we would like to reply to the comment and clarify some facts in arXiv:1407.5194.
\end{abstract}

%\pacs{02.30.Ik,03.75.Lm, 42.81.Dp}
 \maketitle

Algebro-geometric solution is a kind of very important solution
in the integrable system. The soliton solution, breather solution and rogue wave solution
can be obtained from the algebro-geometric solution through special limit.
Therefore, we agree that almost all the ``new soliton solutions"
of the vector NLS equation are contained in the general formula in \cite{1} as
particular cases. However, it is still meaningful to find out the different type nonlinear localized wave solutions explicitly. Based on the explicit expressions, it is convenient to study on their physical property and potential applications. For instance, Akhmediev breather and rational solution of scalar NLS equation has been used to direct experiments to excite rogue wave in many physical systems \cite{Kibler, Chabchoub}. There have been
lots of works about the reduction of algebro-geometric solution, such as the recent work \cite{kalla}. In the paper \cite{2}, we presented an explicit expression for many types localized wave solutions in a unified form in a different way. The derivation method and solution form are both different from the ones in \cite{1}. Furthermore, we discuss the classification
of the general soliton formula on the nonzero background based on the dynamical behavior.
Especially, the conditions for breather, dark soliton and rogue wave solution for mCNLSE are
given in detail. These results are also quite distinctive from the results before.

We fell sorry about the misunderstanding induced by some representations in the manuscript. We will revise them seriously. We are grateful to the author for his comment. Here, it is still need to reply to the explicit comments one by one.

\begin{enumerate}
  \item
The Theorem 5 in \cite{2} give the $N$-fold Darboux transformation with the loop group form \cite{Terng}. However, no theorem was found to describe the Darboux transformation of AKNS system in reference \cite{1}, since they used the generalized algebraic geometry (or finite gap) scheme, and the derivation methods are distinctive well.
On other hand, we give a proof to the non-singularity of $N$-localized wave solution in the theorem 6. Indeed, this theorem is consistent with theorem 2 and theorem 3
in reference \cite{1}. However, the solution is represented in a different form. In reference \cite{1}, the determinant is $2N\times 2N$ for breather solution,  and the determinant is $N\times N$ for
dark soliton solution. In our work \cite{2}, the determinant is written in $N\times N$ uniformly for these nonlinear wave solutions and even rogue wave solution. $\square$

\item
The equations (16) in \cite{2} are merely to look for compact formula
of breather solution or dark soliton solution. For instance, we have
\begin{equation}\label{eq1}
   2\lambda-\chi-\frac{c_1^2}{\chi+a_1}+\frac{c_2^2}{\chi+a_2}=0,
\end{equation}
taking conjugation to above equation, it follows that
\begin{equation}\label{eqc1}
   2\bar{\lambda}-\bar{\chi}-\frac{c_1^2}{\bar{\chi}+a_1}+\frac{c_2^2}{\bar{\chi}+a_2}=0,
\end{equation}
where overbar represents the complex conjugation.
If we use \eqref{eq1} minus \eqref{eqc1}, it follows that
\begin{equation*}
     2(\lambda-\bar{\lambda})+\bar{\chi}-\chi-\frac{c_1^2(\bar{\chi}-\chi)}{|\chi+a_1|^2}+\frac{c_2^2(\bar{\chi}-\chi)}{|\chi+a_2|^2}=0.
\end{equation*}
Moreover, it follows that
\begin{equation}\label{identity1}
     1-\frac{c_1^2}{|\chi+a_1|^2}+\frac{c_2^2}{|\chi+a_2|^2}=\frac{2(\lambda-\bar{\lambda})}{\chi-\bar{\chi}}.
\end{equation}
In a similar way, we can obtain the other algebraic identity. Maybe the above identity is obtained with a trick way. To
obtain a generalized trick, we can use the following relation:
\begin{equation}\label{rela1}
    \Phi^{\dag}(\mu)J\Phi(\nu)=i(\nu-\bar{\mu})\int \Phi^{\dag}(\mu)J\sigma_3\Phi(\nu) dx,
\end{equation}
where
\begin{equation*}
\begin{split}
   [\Phi(\nu)]_x & =(i\nu \sigma_3+Q)\Phi(\nu), \\
   -[\Phi^{\dag}(\mu)J]_x&=\Phi^{\dag}(\mu)J(i\bar{\mu} \sigma_3+Q).
\end{split}
\end{equation*}
Substituting
\begin{equation*}
    \Phi(\lambda)=\begin{bmatrix}
           1 \\
           e^{i\theta_1}\alpha_1 \\
           e^{i\theta_1}\alpha_2 \\
         \end{bmatrix}e^{i(\chi-\lambda)x},\,\, \alpha_1=\frac{c_1}{\chi+a_1},\,\, \alpha_2=\frac{c_2}{\chi+a_2},
\end{equation*}
into relation \eqref{rela1}, we have
\begin{equation*}
    1-|\alpha_1|^2+|\alpha_2|^2=\frac{\lambda-\bar{\lambda}}{(\chi-\lambda)-(\bar{\chi}-\bar{\lambda})}(1+|\alpha_1|^2-|\alpha_2|^2).
\end{equation*}
It follows that we have \eqref{identity1}:
\begin{equation*}
   1-|\alpha_1|^2+|\alpha_2|^2=\frac{2
    (\lambda-\bar{\lambda})}{\chi-\bar{\chi}}.
\end{equation*}

In reference \cite{4}, the author uses the eigenvalue of characteristic equation as the free parameter. The method presented there
can be used to solve a quantic equation instead of the cubic equation for the two component coupled NLS equations. This is a good mathematical technique to deal with this problem.
Indeed, our work do not hamper the using of this technique. But on the other hand, when the degree of algebraic equation is more than 5, there is no general
roots formula. Usually, we solve the algebraic solution by numeric method. Even though the cubic equation, we use the numeric method instead of Cardano's formula. $\square$

\item
On the other hand, we should stress that our method presented in \cite{2} can be readily generalized to the more general case. For instance, the matrix NLS
equations, $N$-component NLS equations and so on.
It was previously deemed that the nonsingular condition of matrix NLS equations is still open \cite{4}. Indeed, since there are lots of parameters for general matrix NLS equations, the classification of nonsingular solution for the matrix NLS equations on the plane wave background is not a simple work.
Moreover, there is no similar algebraic identity \eqref{identity1} for the matrix NLS equations.
But our method can be applied to solve this problem for a certainty.  We give the following example to illustrate this explicitly.

\noindent\textbf{Example: The $2\times 2$ matrix NLS equations.}

Here we use the notation in reference \cite{4}. We consider the matrix in (2.22a) of reference \cite{4}
\begin{equation*}
U_0=\begin{bmatrix}
  -\mu I_l & A \\
  -\Sigma A^{\dag} \Omega & \mu I_m +\Gamma \\
\end{bmatrix}.
\end{equation*}
To illustrate our method clearly, we take $l=m=2$, $\Sigma=\Omega=\mathrm{diag}(1,-1)$ and
\begin{equation*}
    A=\begin{bmatrix}
        A_1 & A_{2} \\
        A_3 & A_4 \\
      \end{bmatrix}.
\end{equation*}
Then we have
\begin{equation*}
    [(\lambda+\mu)-U_0]\begin{bmatrix}
                         1 \\
                         \beta \\
                         \frac{-\bar{A}_1+\beta \bar{A}_3}{\lambda-\gamma_1} \\
                         \frac{\bar{A}_2-\beta \bar{A}_4}{\lambda-\gamma_2} \\
                       \end{bmatrix}
    =\begin{bmatrix}
          2\mu+\lambda & 0 & -A_1 & -A_2 \\
          0 & 2\mu+\lambda & -A_3 & -A_4 \\
          \bar{A}_1 & -\bar{A}_3 & \lambda-\gamma_1 & 0 \\
           -\bar{A}_2 & \bar{A}_4 & 0 & \lambda-\gamma_2 \\
        \end{bmatrix}\begin{bmatrix}
                         1 \\
                         \beta \\
                         \frac{-\bar{A}_1+\beta \bar{A}_3}{\lambda-\gamma_1} \\
                         \frac{\bar{A}_2-\beta \bar{A}_4}{\lambda-\gamma_2} \\
                       \end{bmatrix}=0,
\end{equation*}
where the parameter $\lambda$ satisfies equation $\det[(\lambda+\mu)-U_0]=0$ (in what following, we merely consider this equation possesses four different roots), the parameter $\beta$ satisfies the following equations:
\begin{equation*}
    \begin{split}
      2\mu+\lambda-A_1\left(\frac{-\bar{A}_1+\beta \bar{A}_3}{\lambda-\gamma_1}\right)-A_2\left(\frac{\bar{A}_2-\beta \bar{A}_4}{\lambda-\gamma_2}\right)=&0,  \\
      (2\mu+\lambda)\beta-A_3\left(\frac{-\bar{A}_1+\beta \bar{A}_3}{\lambda-\gamma_1}\right)-A_4\left(\frac{\bar{A}_2-\beta \bar{A}_4}{\lambda-\gamma_2}\right)=&0.
    \end{split}
\end{equation*}
Then we can obtain a vector solution with the following representation:
\begin{equation}\label{phi}
    \Phi(\mu)=D\begin{bmatrix}
                         1 \\
                         \beta \\
                         \frac{-\bar{A}_1+\beta \bar{A}_3}{\lambda+\gamma_1} \\
                         \frac{\bar{A}_2-\beta \bar{A}_4}{\lambda+\gamma_2} \\
                       \end{bmatrix}e^{i(\lambda+\mu)x+i(-\lambda^2+2\mu^2)t},\,\, D=\begin{bmatrix}
           P_1e^{-2itAB} & O \\
           O & iP_2^{-1}e^{-ix\Gamma+it\Gamma^2} \\
         \end{bmatrix}.
\end{equation}
By Darboux transformation with loop group version \cite{2,Terng}, the above matrix NLS system possesses the following
Darboux transformation:
\begin{equation*}
    T=I+\frac{\bar{\mu}_1-\mu_1}{\mu-\bar{\mu}_1}\frac{|y_1\rangle\langle y_1|J}{\langle y_1|J|y_1\rangle}, \,\, J=\mathrm{diag}(\Omega,-\Sigma),
\end{equation*}
where $|y_1\rangle=v(x,t)\Phi_1$, $v_1(x,t)$ is a nonzero analytical function and $\Phi_1$ is a special solution $\Phi$ \eqref{phi} for Lax pair equation with $\mu=\mu_1$.

To obtain the nonsingular solution for matrix NLS system is equivalent to find the regular Darboux matrix. So we merely need to analyze the quadric form
$\langle y_1|J|y_1\rangle$.
For the $2\times 2$ matrix NLS system, we take $v_1(x,t)=e^{-i\mu_1(x+2\mu_1t)}.$ It follows that
\begin{equation*}
    |y_1\rangle=DM_1Y_1,
\end{equation*}
where
\begin{equation*}
    M_1=\begin{bmatrix}
    1&1&1&1 \\
    \beta_1&\beta_2&\beta_3&\beta_4 \\
    \frac{-\bar{A}_1+\beta_1 \bar{A}_3}{\lambda_1+\gamma_1}&\frac{-\bar{A}_1+\beta_2 \bar{A}_3}{\lambda_2+\gamma_1}
    &\frac{-\bar{A}_1+\beta_3 \bar{A}_3}{\lambda_3+\gamma_1}&\frac{-\bar{A}_1+\beta_4 \bar{A}_3}{\lambda_4+\gamma_1} \\
    \frac{\bar{A}_2-\beta_4 \bar{A}}{\lambda_1+\gamma_2}&\frac{\bar{A}_2-\beta_2 \bar{A}_4}{\lambda_2+\gamma_2}
    &\frac{\bar{A}_2-\beta_3 \bar{A}_4}{\lambda_3+\gamma_2}&\frac{\bar{A}_2-\beta_4 \bar{A}_4}{\lambda_4+\gamma_2} \\
    \end{bmatrix},\,\,Y_1=\begin{bmatrix}
     d_1e^{i\lambda_1(x-\lambda_1t)} \\
     d_2e^{i\lambda_2(x-\lambda_2t)} \\
     d_3e^{i\lambda_3(x-\lambda_3t)} \\
     d_4e^{i\lambda_4(x-\lambda_4t)} \\
\end{bmatrix},
\end{equation*}
$\lambda_i$ $(i=1,2,3,4)$ are four different roots for equation $\det[(\lambda+\mu_1)-U_0]=0$, $d_i$ $(i=1,2,3,4)$ are complex constants. With the aid of similar relation as \eqref{rela1}, we have
\begin{equation}\label{matrix1}
   N_1\equiv M_1^{\dag}JM_1=2(\bar{\mu}_1-\mu_1)\begin{bmatrix}
                                      \frac{1-|\beta_1|^2}{\left(\lambda_1-\bar{\lambda}_1\right)} & \frac{1-\beta_2\bar{\beta}_1}
                                      {\left(\lambda_2-\bar{\lambda}_1\right)} & \frac{1-\beta_3\bar{\beta}_1}
                                      {\left(\lambda_3-\bar{\lambda}_1\right)} & \frac{1-\beta_4\bar{\beta}_1}
                                      {\left(\lambda_4-\bar{\lambda}_1\right)} \\[8pt]
                                     \frac{1-\beta_1\bar{\beta}_2}{\left(\lambda_1-\bar{\lambda}_2\right)} & \frac{1-|\beta_2|^2}
                                      {\left(\lambda_2-\bar{\lambda}_2\right)} & \frac{1-\beta_3\bar{\beta}_2}
                                      {\left(\lambda_3-\bar{\lambda}_2\right)} & \frac{1-\beta_4\bar{\beta}_2}
                                      {\left(\lambda_4-\bar{\lambda}_2\right)} \\[8pt]
                                      \frac{1-\beta_1\bar{\beta}_3}{\left(\lambda_1-\bar{\lambda}_3\right)} & \frac{1-\beta_2\bar{\beta}_3}
                                      {\left(\lambda_2-\bar{\lambda}_3\right)} & \frac{1-|\beta_3|^2}
                                      {\left(\lambda_3-\bar{\lambda}_3\right)} & \frac{1-\beta_4\bar{\beta}_3}
                                      {\left(\lambda_4-\bar{\lambda}_3\right)} \\[8pt]
                                       \frac{1-\beta_1\bar{\beta}_4}{\left(\lambda_1-\bar{\lambda}_4\right)} & \frac{1-\beta_2\bar{\beta}_4}
                                      {\left(\lambda_2-\bar{\lambda}_4\right)} & \frac{1-\beta_3\bar{\beta}_4}
                                      {\left(\lambda_3-\bar{\lambda}_4\right)} & \frac{1-|\beta_4|^2}
                                      {\left(\lambda_4-\bar{\lambda}_4\right)} \\
                                    \end{bmatrix},
\end{equation}
the parameters are arranged with $$\frac{(|\beta_1|^2-1)}{\mathrm{Im}(\lambda_1)}>\frac{(|\beta_2|^2-1)}{\mathrm{Im}(\lambda_2)}
>\frac{(|\beta_3|^2-1)}{\mathrm{Im}(\lambda_3)}>\frac{(|\beta_4|^2-1)}{\mathrm{Im}(\lambda_4)},\,\, \mathrm{Im}(\mu_1)>0.$$
Since the matrix $N_1$ is congruent with matrix $J=\mathrm{diag}(1,-1,-1,1)$, there are two positive eigenvalues and two negative eigenvalues of matrix $N_1$.
It is naturally to guess that there is one $2\times 2$ positive definite submatrix and one $2\times 2$ negative definite submatrix.
Then we need to look for the positive definite or negative definite submatrix of $N_1$. With the aid of the the positive definite or negative definite submatrices, we could construct the nonsingular solutions on the plane wave background for matrix NLS system by Darboux transformation. To describe the steps
more clearly, we give the following explicit steps:

\textbf{Step 1}: Take $\mu_1=i$, $A_1=2$, $A_2=A_3=A_4=1.$ With the help of Maple soft, the equation $\det[(\lambda+\mu_1)-U_0]=0$ can be solved:
\begin{equation*}
    \begin{split}
      \lambda_1=& -1.01676335958488+0.0648551776223440i, \\
      \lambda_2=& 0.295665613559558-2.88199079701791i,  \\
      \lambda_3=& 0.796680698579354+0.949992083385895i, \\
      \lambda_4=& -0.0755829525540301-2.13285646399033i.
    \end{split}
\end{equation*}
It follows that
\begin{equation*}
    \begin{split}
       \beta_1= &1.11047084620632+0.160225055216483i, \\
       \beta_2= &0.438390662257818+0.0740910597644230i, \\
       \beta_3= &0.494184441353848-0.0404379647066445i,  \\
       \beta_4= &2.09429310597600-0.417054115939492i,
    \end{split}
\end{equation*}
and
\begin{equation*}
    \begin{split}
     \frac{2\mathrm{Im}(\mu_1)}{\mathrm{Im}(\lambda_1)}(|\beta_1|^2-1)=&7.9814003471063, \\
     \frac{2\mathrm{Im}(\mu_1)}{\mathrm{Im}(\lambda_2)}(|\beta_2|^2-1)=&0.556784666306587,  \\
      \frac{2\mathrm{Im}(\mu_1)}{\mathrm{Im}(\lambda_3)}(|\beta_3|^2-1)=&-1.58769009157696,  \\
      \frac{2\mathrm{Im}(\mu_1)}{\mathrm{Im}(\lambda_4)}(|\beta_4|^2-1)=&-3.33824409608918.
    \end{split}
\end{equation*}
\textbf{Step 2}:
By the Courant minimax principle, which is given in the appendix, there is at most one $2\times 2$ positive definite submatrix and one $2\times2$ negative
definite submatrix.
We find that the matrix
\begin{equation*}
    N_{1+}= \left[ \begin {array}{cc}  7.981400347&
 0.5783206801- 0.2865122949\,i\\ \noalign{\medskip}
 0.5783206801+ 0.2865122949\,i& 0.5567846663\end {array} \right]
\end{equation*}
is positive definite;
and the matrix
\begin{equation*}
    N_{1-}= \left[ \begin {array}{cc} - 1.587690091&-
 0.3096538015+ 0.1822280443\,i\\ \noalign{\medskip}-
 0.3096538015- 0.1822280443\,i&- 3.338244096\end {array} \right]
\end{equation*}
is negative definite.

\textbf{Step 3:}
Then we can choose two special solutions to construct the nonsingular solution of matrix NLS equations:
\begin{itemize}
  \item The first case is $d_1\neq0$, $d_2\neq0$ and $d_3=d_4=0,$
  \item The second case is $d_3\neq0$, $d_4\neq0$ and $d_1=d_2=0.$
\end{itemize}

The rest things is substituting the special vector solutions into solution formula to obtain the exact solution for matrix NLS equations. This is all the steps. $\square$
\end{enumerate}

Finally, we clarify some misunderstandings between us and the author of comment.
The previous sentence ``since Tsuchida's work did not consider the limit technique, he can
not obtain the rogue wave solution." is indeed not proper. It really means that  ``he has not
presented the rogue wave solution." Maybe there are still some other improper or unprecise sentences in the paper for arXiv.  We are sorry for improper presentations on the compare with Prof. Tsuchida's work. We will revise it seriously with the help of colleagues who are more professional in English.

It is emphasized that the method presented in \cite{2} can be extended to derive nonlinear wave solutions of many different type nonlinear partial equations. We merely solve the mixed coupled NLS equations to demonstrate the method. The related papers for de-focusing or focusing NLS equations were cited properly in the paper.

\section*{Appendix}
\begin{thm}[\cite{courant},Courant minimax principle]
Let $A$ be $n\times n$ a Hermite matrix with eigenvalues $\lambda_1\leq \lambda_2\leq \cdots \leq \lambda_n$, and the diagonal elements
of matrix $A$ are $a_{11},a_{22},\cdots,a_{nn}$, then
\begin{equation*}
    \lambda_1\leq \max \limits_{i} \{a_{ii}\}, \min \limits_{i} \{a_{ii}\}  \leq \lambda_n
\end{equation*}
\end{thm}

\section*{Acknowledgments}
The authors are grateful to Professor Jingsong He for providing us the reference \cite{1}.

\end{document}